\documentclass[aps,prd,amssymb,eqsecnum]{revtex4}

\begin{document}



%
%

\title{Adiabatic radiation reaction to the orbits 
in Kerr Spacetime}

\author{Norichika Sago$^1$, Takahiro Tanaka$^2$,
Wataru Hikida$^3$ and Hiroyuki Nakano$^4$}

\affiliation{
$^1$
Department of Earth and Space Science,~Graduate School of Science,
Osaka University, Toyonaka, Osaka 560-0043, Japan,\\
$^2$
Department of Physics,~Graduate School of Science, Kyoto University,
Kyoto 606-8502,~Japan,\\
$^3$
Yukawa Institute for Theoretical Physics, Kyoto University,
Kyoto 606-8502,~Japan,\\
$^4$Department of Mathematics and Physics, Graduate School of Science,
Osaka City University, Osaka 558-8585, Japan
}


\begin{abstract}
Geodesic motion of a point particle in Kerr geometry has 
three constants of motion, energy $E$,
azimuthal angular momentum $L$, and Carter constant $Q$.
Under the adiabatic approximation, 
radiation reaction effect is 
characterized by the time evolution of these constants.
In this letter 
we show that the scheme to evaluate them
can be dramatically simplified. 
\end{abstract}


\maketitle

\section{Introduction}
A supermassive black hole (SMBH) accompanied by a compact object
(CO) is among
the most promising candidates for gravitational wave sources.
This system may provide us the best opportunity for  
testing general relativity in the strong gravity regime. 
For this purpose, however, we need an accurate theoretical 
prediction of its waveform. 

To investigate gravitational waves from SMBH-CO binary system,
we use the black hole perturbation method. 
The background geometry is Kerr spacetime, 
and CO is described by a point particle.
In the lowest order in mass ratio, the particle moves
along the background geodesic. In the next order its orbit
deviates from the geodesic due to radiation reaction effects. 

In the Schwarzschild background, the particle's orbit can be
characterized solely by energy $E$ and azimuthal angular
momentum $L$. We can evaluate the orbital evolution from 
the change rates of energy and azimuthal angular momentum, 
$d{E}/dt$ and $d{L}/dt$. 
In the adiabatic approximation
the energy and the angular momentum that a particle loses 
are equated with the ones radiated to the infinity or into the black 
hole horizon as gravitational waves since there are conservation lows 
for $E$ and $L$ including the gravitational field.  
In this manner we can determine $dE/dt$ and $dL/dt$
from the asymptotic behavior of the radiated gravitational waves 
in the Schwarzschild case.

On the other hand, Carter constant $Q$ is also necessary
to specify a geodesic in Kerr background. 
Since we do not have a conservation law corresponding to 
Carter constant, the change rate of $Q$ is not 
directly related to the asymptotic gravitational waves. 
Instead, we need to directly calculate the self-force acting on 
the particle~\cite{Ori95}.
Though the prescription to calculate the self-force is formally
established~\cite{MST97,QW97}, performing 
explicit calculation is not so straight forward.

Gal'tsov~\cite{Gal'tsov82} employed the \emph{radiative} part of the
metric perturbation, which was introduced earlier by 
Dirac~\cite{Dirac38}, 
to calculate $d{E}/dt$ and $d{L}/dt$. 
He showed that this scheme correctly reproduces the standard results 
obtained by using the conservation laws. 
Recently, Mino gave a justification for applying the same scheme
to $dQ/dt$ for bound orbits~\cite{Mino03}.
(See also Ref.~\cite{Hughes05}.)
However, actual implementation of $dQ/dt$ calculation again 
is not so straight forward. 
In this letter we derive a rather simple new 
formula for the 
adiabatic evolution of Carter constant.

\section{Background}
We consider the background Kerr spacetime in the 
Boyer-Lindquist coordinates: 
$
ds^2=-(1-2Mr/\Sigma)dt^2
-(4Mar\sin^2\theta/\Sigma)dtd\varphi
+(\Sigma/\Delta)dr^2+\Sigma d\theta^2
+(r^2+a^2+2Ma^2r\sin^2\theta/\Sigma)
\sin^2\theta d\varphi^2, 
$
where
$
\Sigma=r^2+a^2\cos^2\theta
$ and $
\Delta=r^2-2Mr+a^2.
$
Here $M$ and $aM$ are the mass and the angular momentum of
the black hole, respectively.
There are two Killing vectors 
$\xi_{(t)}^{\mu}:
=(\partial_t)^\mu
$ and $
 \xi_{(\varphi)}^{\mu}:
=(\partial_\varphi)^\mu$ .
In addition, Kerr spacetime possesses the Killing tensor,
$
K_{\mu\nu}:
=2\Sigma l_{(\mu}n_{\nu)}+r^2g_{\mu\nu},
$
where the parentheses
denote symmetrization on the indices enclosed, and
$
l^{\mu}:=
\left({r^2+a^2},\Delta,0,a \right)/\Delta
$
and
$
n^{\mu}:=
\left(r^2+a^2,-\Delta,0,a\right)/2\Sigma
$
are two radial null vectors.
Killing tensor satisfies the equation $K_{(\mu\nu;\rho)}=0$,

We consider motion of a point particle, 
$z^{\alpha}(\tau) =
(t_z(\tau),r_z(\tau),\theta_z(\tau),\phi_z(\tau))$. 
Here $\tau$ is the proper time along the orbit.
For geodesic motion, there are 
three constants of motion, 
\begin{eqnarray}
E&:=&-u^{\alpha}\xi_{\alpha}^{(t)}=
\left(1-\frac{2Mr_z}{\Sigma}\right)u^t
+\frac{2Mar_z\sin^2\theta_z}{\Sigma}u^{\varphi}, \label{eq:Energy} \\
L&:=&u^{\alpha}\xi_{\alpha}^{(\varphi)}=
-\frac{2Mar_z\sin^2\theta_z}{\Sigma}u^t
+\frac{(r_z^2+a^2)^2-\Delta a^2\sin^2\theta_z}{\Sigma}
\sin^2\theta_z u^\varphi, \label{eq:Momentum} \\
Q&:=&K_{\alpha\beta}u^{\alpha}u^{\beta}
=\frac{(L-aE\sin^2\theta_z)^2}{\sin^2\theta_z}
+a^2\cos^2\theta_z+\Sigma^2 (u^\theta)^2, \label{eq:Carter}
\end{eqnarray}
where $u^{\alpha}:
=dz^{\alpha}/d\tau$.
In addition, we define another notation for the Carter constant, 
$C :=
 Q-(aE-L)^2$. 
For orbits on the equatorial plane $C$ vanishes.

\section{Geodesic Motion in Kerr Spacetime}
Introducing a new parameter $\lambda$ defined by 
$d\lambda=d\tau/\Sigma$,
the geodesic equations become 
\begin{eqnarray}
&& \left(\frac{dr_z}{d\lambda}\right)^2 =
R(r_z), \qquad \left(\frac{d\cos\theta_z}{d\lambda}\right)^2 =
\Theta(\cos\theta_z),\\
&& \frac{dt_z}{d\lambda} = 
-a(aE\sin^2\theta_z-L)
+\frac{r_z^2+a^2}{\Delta}P(r_z),~
\frac{d\varphi_z}{d\lambda} =
-aE+\frac{L}{\sin^2\theta_z}
+\frac{a}{\Delta}P(r_z), \label{eq:eom_phi}
\end{eqnarray}
where
$
P(r)=E(r^2+a^2)-aL,
R(r)=[P(r)]^2-\Delta[r^2+Q]
$ and $
\Theta(\cos\theta)=
C - (C+a^2(1-E^2)+L^2)\cos^2\theta+
 a^2(1-E^2)\cos^4\theta.
$
It should be noted that the equation for the $r$-component and
the one for the $\theta$-component are decoubled by using  
$\lambda$.
Both $R(r_z)$ and $\Theta(\cos\theta_z)$ are quartic functions  
of their arguments. Hence both solutions are 
given by elliptic functions. 
For bound orbits, we can systematically expand $r_z$ and
$\cos\theta_z$ in Fourier series.

The other two equations (\ref{eq:eom_phi}) 
are integrated as 
\begin{eqnarray}
t_z(\lambda)&=&t^{(r)}(\lambda)+
   t^{(\theta)}(\lambda)+
   \left\langle {dt_z\over d\lambda}\right\rangle \lambda,\cr
\varphi_z(\lambda)&=&\varphi^{(r)}(\lambda)+
   \varphi^{(\theta)}(\lambda)+
   \left\langle {d\varphi_z\over d\lambda}\right\rangle \lambda, 
\end{eqnarray}
where $\langle \cdots \rangle$ means time average along the  
geodesic. 
$
t^{(r)}(\lambda) := \int d\lambda 
      \{(r_z^2+a^2)P(r_z) /\Delta
$
$
      -\langle(r_z^2+a^2)P(r_z) /\Delta\rangle\}
$
and 
$
t^{(\theta)}(\lambda) := -\int d\lambda 
      \{a(aE\sin^2\theta_z-L)-
$$
            \langle a(aE\sin^2\theta_z-L)\rangle\},  
$
are periodic functions with periods $2\pi \Omega_r$ 
and $2\pi \Omega_\theta$, respectively. 
Functions $\varphi^{(r)}$ and
$\varphi^{(\theta)}$ are also defined in a similar way.

\section{Adiabatic evolution of constants of motion}
In Ref.~\cite{Mino03}, it was shown that the adiabatic radiation 
reaction to the constants of motion $I^i=\left\{E,L,Q\right\}$ 
can be evaluated by 
\begin{equation}
 \left\langle \frac{dI^i}{d\lambda} \right\rangle =
\lim_{T\to\infty}\frac{1}{2T}\int_{-T}^{T}d\lambda\,\Sigma 
 {\partial I^i\over\partial u^\alpha}
 {f}^{\alpha}[h_{\mu\nu}^{\rm rad}],
\label{eq:meanQdot}
\end{equation}
where $h_{\mu\nu}^{{\rm rad}}$ is the radiative part of the
metric perturbation defined by half retarded field minus half 
advanced field, i.e.,  
$
h_{\mu\nu}^{\rm rad}:=
(h_{\mu\nu}^{\rm ret}-h_{\mu\nu}^{\rm adv})/2. 
$
Radiative field is a solution of source free vacuum 
Einstein equation. The singular parts contained in both retarded 
and advanced fields cancel out. 
Therefore we can avoid the tedious issue of regularizing 
self-force. ${f}^{\alpha}$ is a differential operator,
\[
f^{\alpha}[h_{\mu\nu}]:=
-\frac{1}{2}(g^{\alpha\beta}+u^{\alpha}u^{\beta})
(h_{\beta\gamma;\delta}+h_{\beta\delta;\gamma}-h_{\gamma\delta;\beta})
u^{\gamma}u^{\delta}.
\]

\subsection{Calculation of $\dot{E}$ and $\dot L$}
It was shown by Gal'tsov~\cite{Gal'tsov82} that 
\begin{eqnarray}
\left\langle {dE\over d\lambda}\right\rangle & = & 
\lim_{T\to\infty}{1\over 2T}\int_{-T}^T d\lambda \Sigma(-\xi_{(t)}^\alpha)
 f_\alpha [h_{\mu\nu}]\cr
 & = &  \lim_{T\to\infty}{1\over 2T}\int_{-T}^T 
   d\lambda \left[
       (-\xi_{(t)}^{\alpha})
       \partial_\alpha \psi(x)\right]_{x\to z(\lambda)}, 
\label{shownbyGaltsov}
\end{eqnarray}
where 
$\psi(x)=\Sigma \tilde u^\mu 
   \tilde u^\nu h_{\mu\nu}/ 2$ and 
$(\tilde u_t,\tilde u_r,\tilde
u_\theta,\tilde u_\varphi):=
(-E,\pm\sqrt{R(r)}/\Delta
,\pm\sqrt{\Theta(\cos\theta)}/\sin\theta ,L)$. 
This vector field $\tilde u_\mu$ 
is an extension of the four velocity of a particle 
in the sense that it satisfies 
$\tilde u_\mu(z(\lambda))=u_\mu(\lambda)$. 
Since in fact $\tilde{u}_{r}$ and $\tilde {u}_\theta$, respectively, 
depend only on $r$ and $\theta$, we can verify the relation, 
$\tilde{u}_{\alpha;\beta}=\tilde{u}_{\beta;\alpha}$.

Furthermore, Gal'tsov has shown\footnote{In Ref.~\cite{Gal'tsov82},
not the in-field but the out-field was used.} that $\psi(x)$
is given by 
\begin{eqnarray}
\psi(x) = i\int {d\omega\over 2\pi\omega} \sum_{\ell,m} 
   \phi^{(in)}_{\omega,\ell,m}(x) 
   \int d\lambda' 
    \overline{\phi^{(in)}_{\omega,\ell,m}(z(\lambda'))},
\label{dEdlambda1}
\end{eqnarray}
where
$
 \phi_{\omega,\ell,m}(x) = \Sigma
       \tilde u_{\mu}(x)\tilde u_{\nu}(x)
      \pi_{\omega,\ell,m}^{\mu\nu}(x) 
 = \Sigma\tilde u^{\mu}(r,\theta)\tilde u^{\nu}(r,\theta) 
    \tau^*_{\mu\nu}(r,\theta) e^{-i\omega t+im\varphi}
    \Delta^2 $
$\,_{-2}R_{\omega,\ell,m}(r)
    \,_{-2}S_{\omega,\ell,m}(\theta). 
$
$\pi_{\omega,\ell,m}$ is an appropriately normalized mode function 
of metric perturbations, which is constructed by applying 
a second rank differential operator $\tau^*_{\mu\nu}$ 
to a mode function of Teukolsky equation
~\cite{Chrzan75}. 
The method to solve 
Teukolsky equation is well established~\cite{Tagoshi}.
We can calculate the contribution from waves absorbed into
a black hole by replacing the in-field to the up-field in
Eq.~(\ref{dEdlambda1}).

We can express $\phi_{\omega,\ell,m}$ as 
\begin{eqnarray}
    \phi_{\omega,\ell,m}(z(\lambda))
 =e^{-i(\omega t_z(\lambda)
         -m \varphi_z(\lambda))} 
       \Phi_{\ell,m}(r_z(\lambda),dr_z/d\lambda(\lambda),
      \theta_z(\lambda),d\theta_z/d\lambda(\lambda)). 
\end{eqnarray}
In the folloing text, we abbreviate 
$dr_z/d\lambda$ and $d\theta_z/d\lambda$ from the arguments 
for brevity. 
Here the exponent contains 
$t^{(r)}(\lambda)$, $t^{(\theta)}(\lambda)$, 
$\varphi^{(r)}(\lambda)$ and 
$\varphi^{(\theta)}(\lambda)$. 
Since $r_z$, $t^{(r)}$ and $\varphi^{(r)}$ 
($\theta_z$, $t^{(\theta)}$ and $\varphi^{(\theta)}$) 
are periodic functions with period $2\pi\Omega_r^{-1}$ 
($2\pi\Omega_\theta^{-1}$), we can expand 
$e^{-i(\omega(t^{(r)}(\lambda_r)+t^{(\theta)}(\lambda_\theta))
   -m(\varphi^{(r)}(\lambda_r)+\varphi^{(\theta)}(\lambda_\theta)
      ))}
\Phi_{\ell,m}(r_z(\lambda_r),\theta_z(\lambda_\theta))$ into 
Fourier series as  
$\langle {dt_z/d\lambda}\rangle
\sum_{n_r,n_\theta}
     Z_{\omega,\ell,m}^{n_r,n_\theta}
     e^{in_r\Omega_r\lambda_r +
    in_\theta\Omega_\theta\lambda_\theta}$
. 
Therefore we obtain
$
 \exp[-i\omega(t^{(r)}(\lambda)+t^{(\theta)}(\lambda))
   +im(\varphi^{(r)}(\lambda)+\varphi^{(\theta)}(\lambda)
      )]
   \Phi_{\ell,m}(r_z(\lambda),\theta_z(\lambda))
  = \left\langle {dt_z / d\lambda}\right\rangle 
   \sum_{n_r,n_\theta} 
    Z_{\omega,\ell,m}^{n_r,n_\theta}
   e^{i(n_r\Omega_r +n_\theta\Omega_\theta)\lambda}.
$
Using this expansion, we obtain 
\begin{eqnarray}
\int d\lambda' 
    \overline{\phi^{(in)}_{\omega,\ell,m}(z(\lambda'))}
 & = & 
    \sum_{n_r,n_\theta} 
     2\pi \delta\left(\omega-\omega_m^{n_r,n_\theta}\right)
      \overline{Z_{\ell,m}^{n_r,n_\theta}},
\label{source}
\end{eqnarray}
where
$\omega_m^{n_r,n_\theta}=
\left\langle {dt_z/ d\lambda}\right\rangle^{-1}   
   \left(m\left\langle d\varphi_z/d\lambda\right\rangle
        +n_r \Omega_r + n_\theta\Omega_\theta\right) $ and 
$Z_{\ell,m}^{n_r,n_\theta}:=Z_{\omega_m^{n_r,n_\theta},
\ell,m}^{n_r,n_\theta}$.
Substituting Eq.~(\ref{source}) into Eq.~(\ref{shownbyGaltsov}) 
with Eq.~(\ref{dEdlambda1}), 
and integrating it over $\omega$, we obtain
\begin{eqnarray}
 \!\!\left\langle{dE\over d\lambda}\right\rangle 
 =     \left\langle{dt_z\over d\lambda}\right\rangle
        \lim_{T\to\infty }
            {-1\over 2T}\int_{-T}^T d\lambda\!\!\!\!\!\!
     \mathop{\sum_{\ell,m,}}_{n_r,n_\theta,n'_r,n'_\theta}\!\!\!\!\!\!
        Z_{\ell,m}^{n_r,n_\theta}
     \overline{Z_{\ell,m}^{n'_r,n'_\theta}}
     e^{i((n_r-n'_r)\Omega_r +(n_\theta-n'_\theta)\Omega_\theta)
    \lambda}. 
\end{eqnarray}
Now integration over $\lambda$ is straight forward. We finally 
end up with the well known formula except for the overall normalization 
depending on the definition of the mode function:
\begin{equation}
 \left\langle{dE\over dt}\right\rangle 
    = -\!\!\!\!\sum_{\ell,m,n_r,n_\theta}\!\!\!\!
       \vert Z_{\ell,m}^{n_r,n_\theta}\vert^2. 
\label{standard}
\end{equation} 
In a similar manner, a formula for the angular momentum loss rate is 
obtained as 
\begin{equation}
\left \langle{dL\over dt}\right\rangle 
    =  -\!\!\!\!\sum_{\ell,m,n_r,n_\theta}\!\!\!\!
       {m\over \omega_m^{n_r,n_\theta}}
       \vert Z_{\ell,m}^{n_r,n_\theta}\vert^2. 
\label{standardLz}
\end{equation}

\subsection{calculation of $\dot{Q}$}
The expression for the radiation reaction to the Carter constant 
can be cast into a form analogous to the 
energy loss rate and the angular momentum loss rate. 
Substituting 
\begin{equation}
f_{\nu}
 =\frac{1}{2}(\partial_{\nu}h_{\alpha\beta})u^{\alpha}u^{\beta}
-\frac{d}{d\tau}(h_{\nu\beta}u^{\beta})
-{1\over 2}u_\nu{d\over d\tau}(u^\alpha u^\beta h_{\alpha\beta}), 
\label{fnu}
\end{equation}
the evolution of Carter constant is given by 
\begin{eqnarray}
   \frac{dQ}{d\tau}
   =  2
  K_{\mu}^{\nu}u^{\mu}f_{\nu}
\approx
2
 \left[
K_{\mu}^{\nu}\tilde{u}^{\mu}
\partial_{\nu}{\psi\over \Sigma}
+h_{\alpha\beta}\tilde{u}^{\alpha}\tilde{u}^{\mu}
(K_{\mu ;\nu}^{\beta}\tilde{u}^{\nu}
-K_{\mu}^{\nu}\tilde{u}^{\beta}_{;\nu})
\right]_{x\to z(\lambda)}.
\label{dqdt}
\end{eqnarray}
Here ``$\approx$'' means that terms which become 
a total derivative or $O(h^2)$ are neglected. 
Furthermore we can show that the second term 
in the above equaiton vanishes by using the facts 
$\tilde{u}_{\alpha;\beta}=\tilde{u}_{\beta;\alpha}$ and 
$K_{(\mu\nu;\rho)}=0$. 
Thus we obtain 
\begin{equation}
\left\langle \frac{dQ}{d\lambda}\right\rangle
=
\lim_{T\to\infty}{1\over T}\int_{-T}^T d\lambda \Sigma
K_{\mu}^{\nu}\tilde{u}^{\mu}
\partial_{\nu}{\psi(x)\over \Sigma}, 
\end{equation}
which is written in 
terms of $\psi(x)$. 
Hence, we find that 
the change rate of Carter constant is obtained by 
replacing $-\xi_{(t)}^{\alpha}\partial_\alpha$ 
in the above expression for $\langle dE/dt\rangle$ given in 
Eq.~(\ref{dEdlambda1}) with
$
 2\Sigma K^\nu_\mu\tilde u^\mu\partial_\nu\Sigma^{-1}. 
$
Then, what we have to evaluate is 
\begin{eqnarray}
\int & d\lambda & \left[\Sigma 
    K^\nu_\mu\tilde u^\mu\partial_\nu{\psi(x)\over\Sigma}
       \right]_{x\to z(\lambda)}\cr
  & = & \int d\lambda  \left[\Sigma 
       \left(
            \Sigma(\ell^\mu\tilde u_\mu n^\nu\partial_\nu 
            +n^\nu\tilde u_\mu \ell^\mu\partial_\nu) 
            +{r^2}\tilde u^\mu\partial_\mu\right)
           {\psi(x)\over\Sigma}\right]_{x\to z(\lambda)}\cr
  & = & \int d\lambda \left[
       \left(-{P(r)\over \Delta}
            ((r^2+a^2)\partial_t+a\partial_{\varphi})
            - {dr_z\over d\lambda}\partial_r\right)
            \psi(x)\right]_{x\to z(\lambda)}~~.
\label{CarterReaction}
\end{eqnarray}
In the last step the last term have been integrated by parts 
using $\tilde u^\mu\partial_\mu=\Sigma^{-1}{d/d\lambda}$, which 
is valid after substitution of $z(\lambda)$. 

\section{Further reduction}
Further simplification is possible. For an arbitrary function of 
$r_z$ and $\theta_z$, we have 
\begin{eqnarray}
 &&\lim_{T\to\infty}{1\over 2T}\int_{-T}^{T} 
    d\lambda \exp\left[{-i\omega_m^{n_r,n_\theta} t_z(\lambda)
         +im\varphi_z(\lambda)} 
         \right]
        f(r_z(\lambda),\theta_z(\lambda))\cr
  && 
 =   {\Omega_r\Omega_\theta\over (2\pi)^2}\int_{0}^{2\pi\Omega_r^{-1}} 
        \hspace{-3mm}d\lambda_r
    \int_{0}^{2\pi\Omega_\theta^{-1}}
       \hspace{-3mm} d\lambda_\theta
     \exp\left[
           -in_r\Omega_r\lambda_r-i\omega_m^{n_r,n_\theta}
              t^{(r)}(\lambda_r)
            +im \varphi^{(r)}(\lambda_r)\right]\cr
  &&\qquad  \times      \exp\left[
          {-in_\theta\Omega_\theta\lambda_\theta
            -i\omega_m^{n_r,n_\theta}
             t^{(\theta)}(\lambda_\theta)
            +im \varphi^{(\theta)}(\lambda_\theta)}\right] 
      f(r_z(\lambda_r),\theta_z(\lambda_\theta)).
\end{eqnarray}
This relation can be easily verified by substituting 
$
\exp[-i\omega_m^{n_r,n_\theta}(t^{(r)}
  (\lambda_r)+t^{(\theta)}(\lambda_\theta))
  +im(\varphi^{(r)}(\lambda_r)+\varphi^{(\theta)}(\lambda_\theta)
      ))]
f(r_z(\lambda_r),\theta_z(\lambda_\theta))=\sum_{n_r,n_\theta}
  f_{n_r,n_\theta}$
$ e^{i(n_r\Omega_r+n_\theta\Omega_\theta)\lambda}$. 
Then, by using the above relation, 
the $\lambda$-integral 
in Eq.~(\ref{CarterReaction}) can be decomposed. The part of 
$\lambda_r$-integral takes the following form, and it can be 
integrated 
by parts as 
\begin{eqnarray}
&& -\int d\lambda_r \left[
        {dr_z\over d\lambda}\partial_r
           \exp\left[
            {-in_r\Omega_r\lambda_r-i\omega t
            +im \varphi}\right]
          f(r,\theta)\right]_{r\to r_z(\lambda_r),t\to t^{(r)}
          (\lambda_r),\varphi\to \varphi^{(r)}(\lambda_r)}
\cr &&\quad
 =-\int d\lambda_r \Biggl\{\left[{d\over d\lambda_r}
         -\left({dt^{(r)}\over d\lambda_r}\right)\partial_t
         -\left({d\varphi^{(r)}\over d\lambda_r}\right)
           \partial_\varphi
            +in_r\Omega_r \right]\cr
 &&\hspace*{1cm}   \times \exp\left[
            {-in_r\Omega_r\lambda_r-i\omega t
            +im \varphi}\right]
         f(r,\theta)\Biggr\}_{r\to r_z(\lambda_r),
        t\to t^{(r)}(\lambda_r),\varphi\to \varphi^{(r)}(\lambda_r)}~~.
\end{eqnarray}
The first term in the last line is a total derivative, and therefore
does not contribute to the average over a long period of time.  
Combining the above results, we obtain
\begin{eqnarray}
\left\langle {dQ\over dt}\right\rangle
  & = &  \left\langle{dt_z\over d\lambda}\right\rangle^{\!\!-1}\!\! 
          \!\! 
          \lim_{T\to\infty}\!\!{-1\over T}\!\!\int_{-T}^T 
        \hspace*{-1mm}
\left(\left[
       \left\langle {(r^2+a^2)P\over \Delta}\right\rangle
                 \partial_t
       +\left\langle {a P\over \Delta}\right\rangle
             \partial_\varphi
             +i n_r\Omega_r \right]
            \psi(x)\right)_{\!\! x\to z(\lambda)} 
  \hspace*{-6mm}d\lambda \cr
  & = &  2\left\langle {(r^2+a^2)P\over \Delta}\right\rangle
         \left\langle{dE\over dt}\right\rangle
        -2\left\langle {a P\over \Delta}\right\rangle
           \left\langle{dL\over dt}\right\rangle
            + 2\!\!\!\!\!\!
           \sum_{\ell,m,n_r,n_\theta}\!\!\!\!\!\! 
           {n_r\Omega_r\over \omega_m^{n_r,n_\theta}} 
                \vert Z_{\ell,m}^{n_r,n_\theta}\vert^2. 
\label{dQdlambda}
\end{eqnarray}
This expression is as easy to evaluate as $\langle dE/dt \rangle$ 
and $\langle dL/dt \rangle$. 
To evaluate the last term, we have 
only to replace $m$ in the expression for 
$\langle dL/dt \rangle$ with 
$n_r \Omega_r$. 

\section{Consistency check}
We know that a circular orbit stays circular under radiation 
reaction~\cite{Kennefick}. This condition becomes
$\langle {dQ/dt}\rangle
 =(2(r^2+a^2)P/ \Delta)
   \langle{dE/dt}\rangle
        -(2a P/\Delta)
$
$
         \langle{dL/dt}\rangle.
$
For circular orbits $Z_{\ell,m}^{n_r,n_\theta}=0$ when $n_r\ne 0$. 
Therefore the last term in Eq.~(\ref{dQdlambda}) vanishes. 
Then Eq.~(\ref{dQdlambda}) agrees with the above condtion that a circular orbit 
stays circular. 

We also know that an orbit on the equatorial plane does not 
leave the equatorial plane by symmetry. This can be 
clearly seen by rewriting the above formula in terms of $C$. 
An identity 
$\langle{dt_z/ d\lambda}\rangle 
\langle{dE/ dt}\rangle 
-\langle{d\varphi_z/ d\lambda}\rangle \langle{dL/ dt}\rangle 
+\sum_{\ell,m,n_r,n_\theta}
   \{(n_r\Omega_r+n_\theta\Omega_\theta)/\omega_m^{n_r,n_\theta}\}
                \vert Z_{\ell,m}^{n_r,n_\theta}\vert^2
=0$
follows from the definition of 
$\omega_m^{n_r,n_\theta}$ with the aid of the expressions for 
$\langle{dE/dt}\rangle$ and $\langle{dL/dt}\rangle$ 
given in Eqs.~(\ref{standard}) and (\ref{standardLz}).
Using this identity, we have
\begin{eqnarray}
\left\langle {dC\over dt}\right\rangle
&=& -2 \left\langle a^2 E \cos^2\theta_z\right\rangle
   \left\langle{dE\over dt}\right\rangle
 -2\left\langle {L \cot^2\theta_z}\right\rangle
            \left\langle{dL\over dt}\right\rangle
 -2\!\!\!\!\!\sum_{\ell,m,n_r,n_\theta}\!\!\!\!\!
      {n_\theta\Omega_\theta\over \omega_m^{n_r,n_\theta}} 
                \vert Z_{\ell,m}^{n_r,n_\theta}\vert^2.
\end{eqnarray}
From this equation it is manifest 
that $\langle dC/dt\rangle=0$ when $\theta=\pi/2$. Notice 
that $Z_{\ell,m}^{n_r,n_\theta}\ne 0$ only for $n_\theta=0$ in 
the case of equatorial orbits.

\begin{acknowledgments}
We would like to thank Y. Mino for useful discussions. 
This work is supported in part by JSPS Research Fellowships for
Young Scientists, Nos.~5919 and 1756, by 
Grant-in-Aid for Scientific Research, No. 16740141 
from Japan Society for the Promotion of Scient (JSPS), 
by Grant-in-Aid for Scientific Research, Nos. 14047212, 
14047214, and by that for the 21st Century COE
by that for the 21st Century COE
"Center for Diversity and Universality in Physics" 
at Kyoto university from the Ministry of
Education, Culture, Sports, Science and Technology (MEXT) of Japan.
\end{acknowledgments}


\begin{thebibliography}{99}
\bibitem{Ori95} A.~Ori, Phys. Lett. A {\rm 202}, 347 (1995);
Phys. Rev. D {\rm 55}, 3444 (1997).

\bibitem{MST97}
Y.~Mino, M.~Sasaki, and T.~Tanaka,
Phys. Rev. D {\rm 55}, 3457 (1997).

\bibitem{QW97}
T.C.~Quinn and R.M.~Wald,
Phys. Rev. D {\rm 56}, 3381 (1997).

\bibitem{Gal'tsov82} D.V.~Gal'tsov,
J.Phys. A \rm{15}, 3737 (1982).

\bibitem{Dirac38}
P.A.M.~Dirac,
Proc. R. Soc. London {\rm A167}, 148 (1938).

\bibitem{Mino03} Y.~Mino,
Phys. Rev. D {\rm 67}, 084027 (2003).

\bibitem{Hughes05}
S.~A.~Hughes, S.~Drasco, E.~E.~Flanagan and J.~Franklin,
arXiv:gr-qc/0504015.



\bibitem{Chrzan75}
P.L.~Chrzanowski,
Phys. Rev. D {\rm 11}, 2042 (1975).


\bibitem{Tagoshi}
M.~Sasaki and H.~Tagoshi,
Living Rev.\ Rel.\  {\bf 6}, 6 (2003).


\bibitem{Kennefick}
  D.~Kennefick and A.~Ori,
  Phys.\ Rev.\ D {\bf 53}, 4319 (1996).
\end{thebibliography}
\end{document}